\documentclass[transmag]{IEEEtran}

\usepackage{amssymb, latexsym, amsmath, amsthm, graphicx, epstopdf, here, color, cite}
\usepackage[utf8]{inputenc} 
\usepackage[T1]{fontenc}

\newtheorem{thm}{Theorem}
\newtheorem{lem}{Lemma}
\newtheorem{prop}{Proposition}
\newtheorem{cor}{Corollary}
\newtheorem*{remark}{Remark}
\newtheorem{defi}{Definition}
\newtheorem*{exam}{Example}

\newcommand{\Prob}{\mathbb{P}}
\newcommand{\Geo}{\mathsf{Geo}}

\newcommand{\GI}{\mathsf{GI}}
\newcommand{\M}{\mathsf{M}}

\IEEEoverridecommandlockouts

\title{On Multiple-Access in Queue-Length Sensitive Systems}

\author{ Daewon Seo, Avhishek Chatterjee, and Lav R.~Varshney
	\thanks{D.~Seo and L.~R.~Varshney were supported in part by the National Science Foundation under grant CCF-1623821 and CCF-1717530. A. Chatterjee was supported in part by the Department of Science and Technology in India under grant DST/INSPIRE/04/2016/001171 and SERB SRG/2019/001809.}
	\thanks{D.~Seo is with the Department of Electrical and Computer Engineering, University of Wisconsin-Madison, Madison, WI 53706 USA (e-mail: dseo24@wisc.edu). A.~Chatterjee is with the Electrical Engineering Department, IIT Madras, Chennai 600036 India (e-mail: avhishek@ee.iitm.ac.in). L.~R.~Varshney is with the Coordinated Science Laboratory, and the Department of Electrical and Computer Engineering, University of Illinois at Urbana-Champaign, Urbana, IL 61801 USA (e-mail: varshney@illinois.edu).}
	\thanks{This work was completed while Seo was at University of Illinois at Urbana-Champaign. This paper was presented in part at the 2018 IEEE International Symposium on Information Theory (ISIT) \cite{SeoCV2018} and is based in part on a thesis submitted in partial fulfillment of the requirements for the degree of Doctor of Philosophy in Electrical and Computer Engineering at University of Illinois at Urbana-Champaign.}

}

\begin{document}

\IEEEtitleabstractindextext{\begin{abstract}
We consider transmission of packets over queue-length sensitive unreliable links, where packets are randomly corrupted through a noisy channel whose transition probabilities are modulated by the queue-length. The goal is to characterize the capacity of this channel. We particularly consider multiple-access systems, where transmitters dispatch encoded symbols over a system that is a superposition of continuous-time $\GI_k/\GI/1$ queues. A server receives and processes symbols in order of arrivals with queue-length dependent noise.

We first determine the capacity of single-user queue-length dependent channels. Further, we characterize the best and worst dispatch processes for $\GI/\M/1$ queues and the best and worst service processes for $\M/\GI/1$ queues. Then, the multiple-access channel capacity is obtained using point processes. When the number of transmitters is large and each arrival process is sparse, the superposition of arrivals approaches a Poisson point process. In characterizing the Poisson approximation, we show that the capacity of the multiple-access system converges to that of a single-user $\M/\GI/1$ queue-length dependent system, and an upper bound on the convergence rate is obtained. This implies that the best and worst server behaviors of single-user $\M/\GI/1$ queues are preserved in the sparse multiple-access case.
\end{abstract}

}

\maketitle

\section{Introduction}
Information systems often need to be resilient to both queuing delays and stochastic noise that corrupt symbols. Yet these aspects are often treated separately in queuing theory and communication theory, respectively. Here, we consider multiple-access settings where a system consists of multiple transmitters that are sending signals. In particular, the sent signals are buffered by a queueing system and queue-lengths modulate noise levels. In this way, communication-theoretic and queuing-theoretic notions of stochasticity are brought together, and one may observe maximizing throughput does not maximize information rate. We find the Shannon capacity of such a queue-length dependent system.

In real-time communication such as live video, voice over Internet protocol (VoIP), machine to machine communication, finance markets \cite{Aldridge2013}, and autonomous vehicles \cite{PapadimitratosFEBC2009}, data packets that are delayed beyond a deadline are as good as undelivered. Consider the scenario where $K$ users in a wireless cell are sending $K$ different VoIP calls via the same base station (BS). Initial packets arrive at the BS and wait at the medium access control (MAC) buffer for proper processing. If the buffer is full, then subsequent packets are often dropped (or erased, a complete loss \cite{SriramL1989}) or their low-quality versions are stored \cite{DraperTW2005} (assuming multi-resolution coding \cite{Goyal2001b}), which leads to information loss and quality deterioration. To remain broad and widely applicable, we consider a generalization of dropping instead of a simple erasure channel---a general noisy channel where the channel quality is dependent on the queue-length seen by arrivals.

With the generalization, we can also think of channel capacity as information processing capability where overloading often lowers job performance of human workers and even machine systems \cite{Schwartz1978, DugdaleEP1999, DerletR2000, Jamal2007}, e.g., due to stress. One motivational setting is driver-assisted autonomous trucks \cite{Higginbotham2019}, where a human driver remotely monitors multiple semi-autonomous trucks and steps in (i.e., processes information) only when the autonomous algorithm cannot handle. Crowdsourcing also belongs to this class of examples. Our work provides a design benchmark for such human-involved information processing systems.

Motivated by applications in multimedia communication and crowdsourcing, we had previously brought some notions of reliability into queueing by establishing the capacity of single-user systems with queue-length dependent service quality \cite{ChatterjeeSV2017}. There, a sequence of coded symbols was sent using an arrival process, processed by an unreliable queueing server, and returned to a destination for decoding. The level of channel noise was a function of the queue-length seen by departure. Then, we investigated the capacity of a single-user queue-length dependent system considering time-slotted (i.e., discrete-time) queues and further optimized a server for $\Geo/\GI/1$ queues or a dispatcher for $\GI/\Geo/1$ queues, under some reliability assumptions.

\begin{figure}[ht]
	\centering
	\includegraphics[width=3.5in]{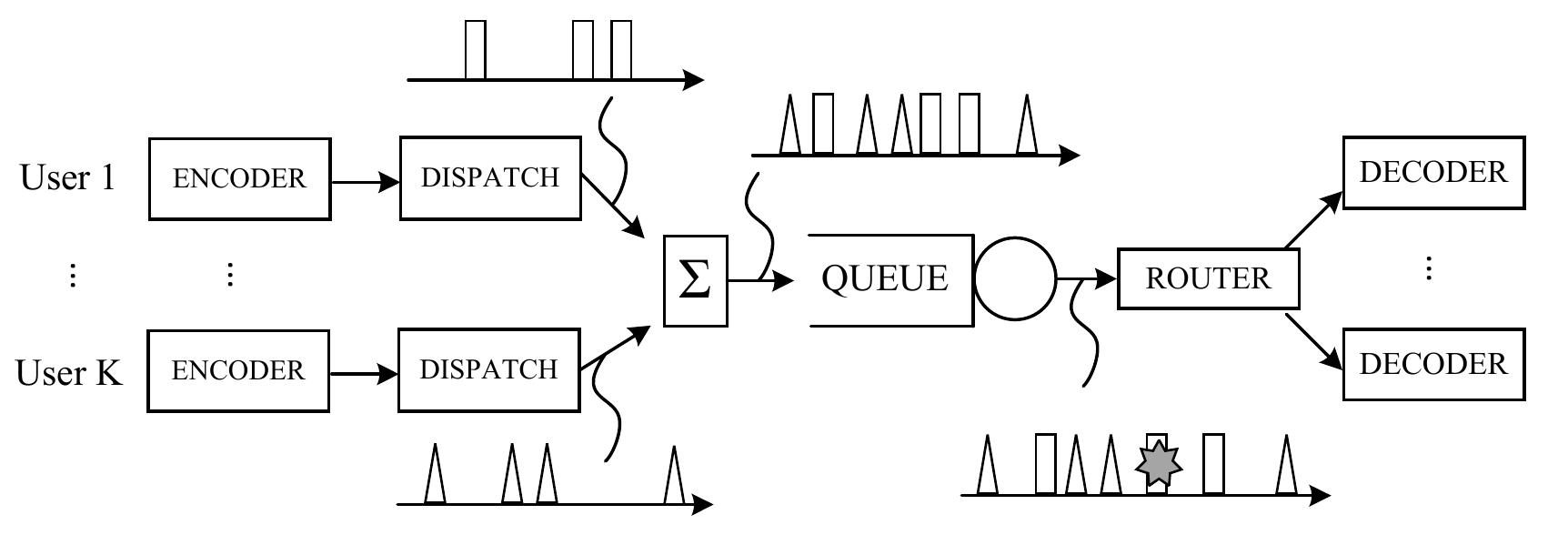}
	\caption{Block diagram of the system: only two out of $K$ point processes are illustrated for brevity. We use $\Sigma$ to denote superposition operation. Encoders map informational bits to encoded symbols, and decoders recover intended informational bits from received noisy symbols.}
	\label{fig:schematic}
\end{figure}

Here, we consider a multiple-access setting using a superposition of multiple arrival processes in continuous-time. Fig.~\ref{fig:schematic} presents the multiple-access setting, where before entering a single central processor, the multiple arrival processes are superposed. Once coded symbols\footnote{Throughout this paper, a \textit{symbol} (common in information theory) and a \textit{job} (or a packet, common in queueing theory) are interchangeable.} arrive at the central queue processor, they are corrupted by a noisy channel that is dependent on the queue-length process seen by arrivals, served in a First Come, First Serve (FCFS) manner, and then sent to the intended receiver. Since our work considers a packet as a single symbol over which we can apply a coding scheme, it finds the fundamental informational limit, i.e., Shannon capacity, of such a system. Packets used in practice are suboptimal in the sense of Shannon capacity since they contain control information that is unrelated to the information to be sent.

Before studying the multiple-access system, we first study the capacity of the continuous-time single-user case and specify the best and worst dispatch processes for a $\GI/\M/1$ queue, and service processes for a $\M/\GI/1$ queue. Then, the capacity expression of the multiple-access setting is given in terms of the stationary distribution of queue-length seen by each user's arrivals. As superposition of non-Poisson arrivals is in general intractable, we consider many-user asymptotics by introducing a random marked point process (RMPP, or simply PP) approach \cite{FrankenKAS1982, DaleyV1998} and applying the superposition convergence to a Poisson point process \cite{Grigelionis1963, Kallenberg2017}. The latter states that the superposition of a large number of \textit{sparse} arrivals is approximately Poisson. Building on this, we prove that the capacity for $\sum_k \GI_k /\GI/1$ queues, where $\Sigma_k$ stands for the superposition, converges to that for single-user $\M/\GI/1$ queues with arrival rate appropriately taken. That is, even though individual arrival processes are non-Poisson, encoding symbols as if it is a single-user $\M/\GI/1$ system with properly scaled arrival rate is asymptotically optimal. It also implies the best and worst services obtained for a single-user $\M/\GI/1$ queue still hold in many-user asymptotics of $\sum_k \GI_k /\GI/1$ queues. 

Recently, there has been significant work on real-time scheduling \cite{HouBK2009, HouK2010, JaramilloS2010} and age of information (AoI) \cite{KaulYG2012, CostaCE2014, KamKNE2015, SunUYKS2017}, which study resource allocation in delay-sensitive networks. The current work is distinct from that body of literature and complements it. Firstly, the metrics of interest in AoI, namely average or maximum age, do not directly capture the information loss of the system. Also, the timely packet throughput metric in real-time scheduling, which measures  fraction of packets delivered on time, does not capture the information loss in a Shannon-theoretic sense. Secondly, as the real-time scheduling and AoI literatures do not consider Shannon capacity, the aspect of coding across packets to increase reliability is absent there. Our approach here yields benchmarks for coding schemes to strive for reliability. Thirdly, though for simplicity we consider only FCFS service, the information capacity of a system can be studied under any service policy. Characterizing the information capacity of a system operated under an efficient real-time scheduling algorithm would help develop codes to be used in conjunction with scheduling.

Like our FCFS model where the server only processes a single job at a time and waiting jobs interfere with the processing job via increased queue-length, the processor sharing model also has multiple-access interference in queuing \cite{TelatarG1995, RajTT2004}. Many-user asymptotics also appear in the many-access channel \cite{ChenCG2017} which studies the capacity of the Gaussian multiple-access channel in terms of the number of users. In continuous-time \cite{AnantharamV1996} and discrete-time \cite{BedekarA1998} timing channels, information is modulated via transmission times in the face of random queuing delays.  No information is included in the symbols themselves.  Contrarily, in our work, there is no timing information and all information is in the symbols themselves.  Note that our multiple-access setting differs from classic multiple-access channels where signals interfere with each other directly \cite{Ahlswede1971, Ahlswede1974}: in our case signal interference only manifests as increased queue length.

The remainder of the paper is organized as follows. Sec.~\ref{sec:model} introduces the queue-length dependent channel and some definitions of point processes. Sec.~\ref{sec:single-user} describes the capacity of the single-user case for continuous-time queueing workloads as well as its best and worst behaviors. Main contributions of this paper are discussed in Sec.~\ref{sec:multi-user}, which states capacity formula for a general $K$-user system and its Poisson approximation when users are sparse. Sec.~\ref{sec:conclusion} concludes the paper.

\section{Preliminaries and System Model} \label{sec:model}
\subsection{Point Processes}
We use a PP approach to queueing systems, enabling us to derive analytic properties. Let us define an RMPP $\Phi=\Phi(t), t \in \mathbb{R}$ as follows.
\begin{defi}
	Let $\mathfrak{B}$ be the Borel $\sigma$-algebra of $\mathbb{R}$. Given a mark space $\mathcal{M}$ and its sigma-algebra $\sigma(\mathcal{M})$, consider a marked counting measure $N(B \times M)$ where $B \in \mathfrak{B}$ and $M \in \sigma(\mathcal{M})$ such that $N(B \times \mathcal{M}) < \infty$ for any bounded $B$. Let $\mathcal{N}, \sigma(\mathcal{N})$ be the set of all such counting measures and its smallest $\sigma$-algebra, respectively. Then, a \textit{random marked point process} (RMPP, or simply a point process (PP)), $\Phi(t)$ is a random element from $(\Omega, \mathcal{F}, P)$ to $(\mathcal{N}, \sigma(\mathcal{N}))$.
\end{defi}

For queueing applications, the mark usually denotes a random service time at the server or the time required to finish each job. Hence, $\mathcal{M} = \mathbb{R}_+ := \{x \in \mathbb{R}| x \ge 0 \}$ and each mark is i.i.d.\ from some distribution $P^S$ since only $\cdot/\GI/1$ queues are considered in this work. In other words, the service time of the $i$th job is $S_i \sim P^S$. Since all randomness from arrival and service times is captured in the RMPP, any queue response such as queue-length or waiting time is a deterministic function of the RMPP.

Two equivalent representations of a PP are especially useful in this paper. Suppose the mark space is empty, i.e., $\mathcal{M} = \emptyset$ for simple exposition, but the following representations can be easily extended to RMPPs with an arbitrary mark space. The first is an inter-arrival time representation, induced by a Dirac-delta train.

Letting $\{ T_i \in \mathbb{R}_+ \}_{i \in \mathbb{Z}}$ be an increasing random sequence, the following representations are all equivalent:
\begin{align*}
	\Phi(t) \Leftrightarrow \sum_{i=-\infty}^{\infty} \delta_{T_i} \Leftrightarrow (\ldots, A_{-1}, A_0, A_1, \ldots),
\end{align*}
where $\Leftrightarrow$ indicates an alternative representation, and $A_i := T_{i}-T_{i-1} \ge 0$. So $T_i$ indicates the time epoch when the $i$th arrival comes. The case when $A_i$ are i.i.d.~is called a \textit{renewal} process, which arises in Sec.~\ref{sec:single-user}.

The other representation is by the counting measure, which is useful especially in Sec.~\ref{sec:multi-user}. Note that
\begin{align*}
	N(B) = \int_{\mathbb{R}} \sum_{i=-\infty}^{\infty} \mathbf{1}_B(s) \delta_{T_i} ds,
\end{align*}
that is, the number of arrivals in $B$, for a bounded $B \in \mathfrak{B}$, uniquely determines $\Phi(t)$. Here, $\mathbf{1}_B = \mathbf{1}_B(s)$ is the indicator function with criterion $\{s \in B\}$, and we write $\mathbf{1}_B \Phi$ to stand for the restricted RMPP on $B$.

A time shift operation is denoted by $\mathcal{T}_{\tau} \Phi(t) = \Phi(t+\tau)$, enabling definitions of stationarity and ergodicity. For a measurable set $Z$, $\mathcal{T}_{\tau} Z := \{ \mathcal{T}_{\tau} \Phi | \Phi \in Z \}$.
\begin{defi}[Stationarity, Def.~1.2.1 \cite{FrankenKAS1982}]
	An RMPP $\Phi$ is \emph{stationary} if the probability measure $P$ is invariant with respect to the time shift $\mathcal{T}_{\tau}$, i.e., for any set $Z \in \sigma(\mathcal{N})$,
	\begin{align*}
		P(\mathcal{T}_{\tau}Z) = P(Z) \mbox{ for all } \tau \in \mathbb{R}.
	\end{align*}
\end{defi}
\begin{defi}[Ergodicity, Def.~1.2.5 \cite{FrankenKAS1982}] \label{def:ergodic_PP}
	A stationary RMPP $\Phi$ (or its probability measure $P$) is \emph{ergodic} if any set $Z \in \sigma(\mathcal{N})$ satisfying $\mathcal{T}_{\tau} Z = Z$ for all $\tau \in \mathbb{R}$ implies either $P(Z) = 0$ or $1$.
\end{defi}

\subsection{System Model} \label{sec:system_model}
Multiple users intend to send messages to respective targeted receivers. To do that, the $k$th user picks an encoded sequence of symbols $X_{(k)}^n$---each symbol is drawn from finite space $\mathcal{X}$---and dispatches it over an independent stationary renewal arrival process with inter-arrival time distribution $P_k^A(t)$. In other words, encoded symbols are sent in sequence on each arrival process. The system only has a single queue so those arrivals are superposed just before entering a queue. The server follows FCFS service discipline with i.i.d.\ service time according to $P^S$. Assume that the waiting room is unlimited.

When the symbol is stored in the queue, it is corrupted to $Y_{(k)}^n \in \mathcal{Y}^n$ randomly, where $\mathcal{Y}$ is also finite. Let $Q$ be the queue-length at the moment just before the symbol's arrival, including the job being serviced. Then, the transition probability $W=W_Q$, i.e., channel, is dependent on $Q$. That is, the channel is $W_Q := P_{Y|X,Q}$. In this sense, we say the system is \emph{queue-length dependent}. We assume that $W_Q \in \mathcal{W}$ for some $|\mathcal{W}| < \infty$. Departing symbols are delivered to the intended receiver. Since symbols are encoded against channel noise, receivers can decode the sequence to recover the original information. We assume that there is a central coordination mechanism that reveals each transmitter's dispatching process to all other transmitters, but not realizations.

As symbols in this work are corrupted at a single queue, the noise depends on other users' transmission. Routing after departing the queue is noiseless. Therefore, we can still call it a multiple-access system, although there are multiple receivers unlike standard multiple-access channels with a single receiver.

We use $\sum(\cdot)$ to denote superposition, so the queue of interest is written as $\sum_k \GI_k/\GI/1$. The queues are assumed always stable, i.e., superposed arrival rate $\lambda$ and service rate $\mu$ satisfy traffic intensity $\rho := \tfrac{\lambda}{\mu} < 1$. We also suppose two technical conditions on arrival and service: 1) arrival and service processes are \textit{simple}, i.e., $P_k^A(t)$ and $P^S(t)$ have no point mass at $t=0$; 2) any or all of  $\{P_k^{A}(t)\}_{k=1}^K$ and $P^S(t)$ are continuous and strictly positive on $\mathbb{R}$.

We assume causal knowledge of arrival and departure realizations, i.e., the encoders do not know them, but the decoders do. Also all $P_k^A$ are available to transmitters, but not their realizations.

\section{Continuous-time Single-user Queue-channel} \label{sec:single-user}
This section investigates the capacity of single-user queue-length dependent channels as in \cite{ChatterjeeSV2017}, but in continuous-time. Before presenting our capacity results, we introduce the inf-information rate, which is used to characterize the capacity of non-i.i.d.\ channels.
\begin{defi}[\cite{VerduH1994, Han2003}]
	The normalized information density is defined as
	\begin{align*}
	\frac{1}{n} i(X^n;Y^n) := \frac{1}{n} \log \frac{\Prob[Y^n|X^n]}{\Prob[Y^n]}.
	\end{align*}
	Then, the \emph{inf-information rate} $\underline{\mathbf{I}}(\mathbf{X};\mathbf{Y})$ is the lim-inf in probability of the normalized information density, i.e., the largest $\alpha \in \mathbb{R} \cup \{\pm \infty\}$ such that for all $\epsilon > 0$,
	\begin{align*}
	\lim_{n \to \infty} \Prob\left[ \frac{1}{n} i(X^n;Y^n) \le \alpha -\epsilon  \right] = 0.
	\end{align*}
\end{defi}
Also, we use $I(P_X, W)$ for standard mutual information \cite{CoverT1991} to clearly indicate its dependency on input distribution $P_X$ and channel $W$.

\subsection{Coding Theorem for $\GI/\GI/1$ Queues}
Consider a simple renewal arrival process $\Phi(t)$ with arrival rate $\lambda$, i.e., the inter-arrival times $\{A_i\}$ are i.i.d.~drawn from $P^A$ with $\lambda = 1/\mathbb{E}[A_1]$. Recall that the channel quality of the $i$th job depends only on the queue-length seen by the $i$th arrival (i.e., just before $i$th arrival), denoted $Q_i$. We first express capacity using the information spectrum method \cite{VerduH1994, Han2003}.

Define the normalized information density conditioned on the queue-length and subsequent conditional inf-information rate,
\begin{align*}
\frac{1}{n} i(X^n;Y^n|Q^n) &= \frac{1}{n} \log \frac{\Prob[Y^n|X^n, Q^n]}{\Prob[Y^n|Q^n]},
\end{align*}
and $\underline{\mathbf{I}}(\mathbf{X};\mathbf{Y}|\mathbf{Q})$ is the lim-inf in probability of $\frac{1}{n} i(X^n;Y^n|Q^n)$. Then, we obtain the following result.

\begin{prop} \label{prop:coding_thm}
	For a simple renewal PP $\Phi(t)$ with rate $\lambda = 1/\mathbb{E}[A_1]$,
	\begin{align}
		C(\Phi) &= \sup_{P_{\mathbf{X}}} \underline{\mathbf{I}}(\mathbf{X};\mathbf{Y}|\mathbf{Q}) ~~~ \textrm{[bits/sym]} \label{eq:cap_formula_multiletter} \\
		&= \sup_{P_{\mathbf{X}}} \lambda \underline{\mathbf{I}}(\mathbf{X};\mathbf{Y}|\mathbf{Q}) ~~~ \textrm{[bits/time]}. \nonumber
	\end{align}
\end{prop}
\begin{IEEEproof}
	The same argument in the proof of \cite[Prop.~1]{ChatterjeeSV2017} holds.
\end{IEEEproof}

\begin{lem} \label{lem:stationary_ergodic_lem}
	For each simple renewal PP $\Phi$, there exists a unique stationary distribution $\pi$ such that if $Q_1$ is drawn from $\pi$, then any $Q_i$ follows $\pi$. Furthermore, for any measurable $f:\mathbb{Z}_+ \mapsto \mathbb{R}_+$, $\frac{1}{n} \sum_{i=1}^n f(Q_i) \to \mathbb{E}_{\pi} [f(Q)]$ as $n \to \infty$ almost surely.
\end{lem}
\begin{IEEEproof}
	Proof follows from standard regenerative cycle argument, e.g., \cite{Asmussen2003}.
\end{IEEEproof}

Combining Prop.~\ref{prop:coding_thm} and Lem.~\ref{lem:stationary_ergodic_lem}, we have a single-letter capacity expression in terms of expectation over $Q$, or equivalently in terms of stationary distribution $\pi(Q)$.
\begin{thm} \label{thm:single_user_cap}
	For $\GI/\GI/1$ queues, the capacity formula \eqref{eq:cap_formula_multiletter} can be further simplified to
	\begin{align}
		C(\Phi) = \sup_{P_X} \mathbb{E}\left[ I(P_X, W_Q)\right] = \sup_{P_X} \sum_{q=0}^{\infty} \pi(q) I(P_X, W_q) \label{eq:ergodic_cap}
	\end{align}
	in bits per job, and
	\begin{align}
		C(\Phi) = \lambda \sup_{P_X} \mathbb{E}\left[ I(P_X, W_Q)\right] = \sup_{P_X} \lambda \sum_{q=0}^{\infty} \pi(q) I(P_X, W_q)
	\end{align}
	in bits per time. Therefore, the capacity over all renewal PPs with stability assumption  $\lambda < \mu$ is
	\begin{align*}
		C = \sup_{\lambda \in (0, \mu)} \sup_{P^A} \sup_{P_X} \lambda \mathbb{E}\left[ I(P_X, W_Q)\right] ~~~ \textrm{[bits/time]}.
	\end{align*}
\end{thm}
\begin{IEEEproof}
	Provided in App.~\ref{sec:app_pf_thm1}.
\end{IEEEproof}

\begin{remark}
In this work, we assume a simple transmitter that does not know arrival and departure realizations, which implies channel state information is unavailable at the encoders. If the channel state information is available without delay (that is, the realizations are known), the capacity formula follows immediately as
	\begin{align}
		C(\Phi) &= \lambda \mathbb{E}\left[ \sup_{P_X} I(P_X, W_Q)\right] ~~~ \textrm{[bits/time]}. \label{eq:ergodic_cap_fullCSI}
	\end{align}
	Thus, we can see that when the capacity-achieving distributions are all identical with some $P_X^*$, such as binary symmetric channels or binary erasure channels, the transmitter simply picks $P_X^*$ even without the channel state information and achieves the same capacity as \eqref{eq:ergodic_cap_fullCSI}. Channel state feedback even without delay does not improve capacity in this case.
\end{remark}

A closed-form expression of $\pi(Q)$ is unknown in general, but is known for some special types of queues. Let us rewrite \eqref{eq:ergodic_cap} for two special types of queues $\GI/\M/1$ and $\M/\GI/1$, and consider per symbol capacity since per time capacity immediately follows by multiplying by $\lambda$.

\begin{cor}[$\GI/\M/1$ queues] \label{thm:G/M/1_queue}
	Let $A^*(\cdot)$ be the Laplace-Stieltjes transform of $P^A(t)$, i.e., $A^*(s) := \int_{0}^{\infty} P^A(t) e^{-st} dt$, and define $\sigma^*$ as the unique solution of $\sigma = A^*(\mu(1-\sigma))$ in $(0,1)$. Then, the capacity of $\GI/\M/1$ queues is given by
	\begin{align*}
		C(\Phi) = \sup_{P_X} \mathbb{E}[I(P_X, W_Q)] ~~~ \textrm{[bits/sym]},
	\end{align*}
	where $\pi(q) = (1-\sigma^*) (\sigma^*)^q$.
\end{cor}
\begin{IEEEproof}
Proof directly follows from standard queueing theory results, e.g., \cite{Kleinrock1975_I}, thus omitted.
\end{IEEEproof}

\begin{cor}[$\M/\GI/1$ queues] \label{thm:M/G/1_queue}
	The capacity of $\M/\GI/1$ queues is given by
	\begin{align*}
		C(\Phi) = \sup_{P_X} \mathbb{E} [I(P_X, W_Q)] ~~~ \textrm{[bits/sym]},
	\end{align*}
	where $\pi(q)$ is obtained from the inverse of probability generating function
	\begin{align*}
		\Pi(z) = \frac{(1-\rho)(1-z)K(z)}{K(z)-z},
	\end{align*}
	and $K(z)$ is the probability generating function of $k_q$ with
	\begin{align*}
		k_q = \int_{0}^{\infty} P^S(t) \frac{e^{-\lambda t} (\lambda t)^q}{q!} dt.
	\end{align*}
\end{cor}
\begin{IEEEproof}
Proof directly follows from standard queueing theory results, e.g., \cite{Kleinrock1975_I}, thus omitted.
\end{IEEEproof}

\begin{figure}[t]
	\centering
	\includegraphics[width=3.5in]{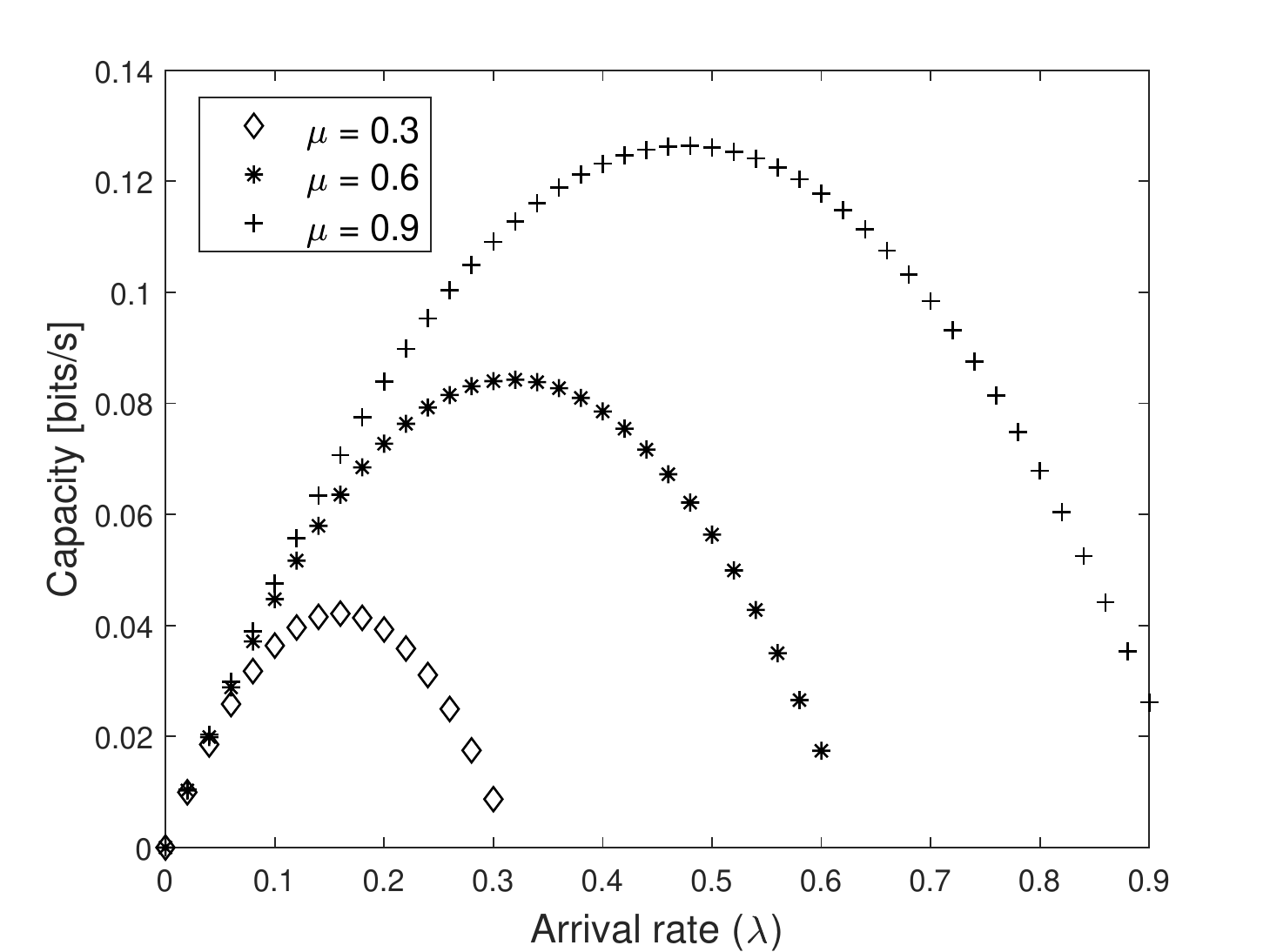}
	\caption{Capacity of $\M/\M/1$ queue (for different service rates) with binary symmetric channel is plotted. $\Prob[X \neq Y] = 0.1$ for $q=0$, $\Prob[X \neq Y] = 0.4$ otherwise. It shows that setting a proper workload maximizes per time capacity.}
	\label{fig:cap_example}
\end{figure}

\begin{exam}
Consider an $\M/\M/1$ queue and a binary symmetric channel with queue-length dependent transition probability $\epsilon_q$. Then, we know that $\pi(q) = (1-\rho)\rho^q$ and Cor.~\ref{thm:G/M/1_queue} shows that
\begin{align*}
C(\lambda) = \lambda \sum_{q=0}^{\infty} \pi(q) (1-H_2(\epsilon_q)) ~~~ \textrm{[bits/time]},
\end{align*}
where $H_2(\cdot)$ is the binary entropy function. Fig.~\ref{fig:cap_example} shows the capacity curves for different service rates. This example implies that there is a nontrivial operating point that maximizes Shannon information rate.
\end{exam}

\subsection{Optimization of Capacity} \label{subsec:optimization_cap}
This subsection considers optimization of the capacities for $\GI/\M/1$ and $\M/\GI/1$ queues given in Cors.~\ref{thm:G/M/1_queue} and \ref{thm:M/G/1_queue}. To do so, we impose two conditions such that
\begin{enumerate}
	\item $P_X^*$ achieves the capacity for all $W_q$.
	
	\item At such $P_X^*$, the system becomes more unreliable as $q$ increases in a step-down manner, i.e., for some $b \in \mathbb{Z}_+$,
	\begin{align*}
		I(P_X^*, W_0) = \cdots = I(P_X^*, W_b) > I(P_X^*, W_{b+1}) = \cdots.
	\end{align*}
\end{enumerate}

Note that condition 1) covers $|\mathbb{F}|$-ary symmetric and $|\mathbb{F}|$-ary erasure channels since $P_X^*$ is uniform. Such channels model random corruption and packet erasure (and then, a dummy packet remains), respectively. In particular, introducing the step-change in noise allows us to find the best and worst server behaviors explicitly. It is natural in applications for the system to be more unreliable as the queue gets longer.

\begin{cor} \label{cor:G/M/1_best}
	Fix arrival rate $\lambda$. For $\GI/\M/1$ queues, the best inter-arrival distribution is deterministic, i.e., $P^A(t)$ only has a unit point mass at $t = \lambda^{-1}$.
\end{cor}
\begin{IEEEproof}
	For the sake of brevity, let $c_b := I(P_X^*,W_b)$ and $c_{b+1} := I(P_X^*,W_{b+1})$. Then, the capacity is written as 
	\begin{align*}
		C(\Phi) &= \sum_{q=0}^{\infty} \pi(q) I(P_X^*, W_q) \\
		&= \sum_{q=0}^{\infty} (1-\sigma^*) (\sigma^*)^q I(P_X^*, W_q) \\
		&= \sum_{q=0}^{b} (1-\sigma^*) (\sigma^*)^q c_b + \sum_{q=b+1}^{\infty} (1-\sigma^*) (\sigma^*)^q c_{b+1} \\
		&= c_b (1-(\sigma^*)^{b+1}) + c_{b+1}(\sigma^*)^{b+1} \\
		&= c_b - (\sigma^*)^{b+1} (c_b - c_{b+1}).
	\end{align*}
	As $c_b > c_{b+1}$, maximizing $C(\Phi)$ with given $\lambda$ is equivalent to minimizing $\sigma^*$. Note that $\sigma^*$ is the unique fixed point of $\sigma = A^*(\mu(1-\sigma))$ and at $\sigma=0$ and $1$,
	\begin{align*}
		\int_{0}^{\infty} P^A(t) e^{-\mu t(1-\sigma)} dt \bigg|_{\sigma=0} &= \int_{0}^{\infty} P^A(t) e^{-\mu t} dt > 0 \\
		\int_{0}^{\infty} P^A(t) e^{-\mu t(1-\sigma)} dt \bigg|_{\sigma=1} &= \int_{0}^{\infty} P^A(t) dt = 1.
	\end{align*}
	Furthermore, $A^*(\mu(1-\sigma))$ is strictly convex in $\sigma$ since
	\begin{align*}
		\frac{\partial}{\partial \sigma} A^*(\mu(1-\sigma)) > 0, ~~~ \frac{\partial^2}{\partial^2 \sigma} A^*(\mu(1-\sigma)) > 0.
	\end{align*}
	Due to Jensen's inequality, we obtain
	\begin{equation} \label{eq:best_arrival}
	\begin{aligned}
		A^*(\mu(1-\sigma)) &= \int_{0}^{\infty} P^A(t) e^{-\mu t(1-\sigma)} dt \\
		&\ge e^{-\mu \mathbb{E}[A] (1-\sigma)} = e^{-\frac{\mu}{\lambda} (1-\sigma)},
	\end{aligned}
	\end{equation}
	where the equality is attained only when $A = \lambda^{-1}$ almost surely. It means that when $P^A$ is deterministic, the curve $A^*(\mu(1-\sigma)) = e^{-\frac{\mu}{\lambda} (1-\sigma)}$ lower bounds all other curves so that achieves the smallest fixed point. Therefore, the deterministic inter-arrival distribution achieves the greatest capacity.
\end{IEEEproof}

\begin{cor} \label{cor:G/M/1_worst}
	Fix arrival rate $\lambda$. For $\GI/\M/1$ queues, cramming inter-arrivals asymptotically minimize the capacity, i.e., $P^A(t;\epsilon,\delta)$ asymptotically achieves the smallest capacity as $\epsilon,\delta \to 0$, where
	\begin{align*}
	P^A(t;\epsilon, \delta) = \begin{cases}
	1-\epsilon & \textrm{if } t=\delta \\
	\epsilon &  \textrm{if } t=\frac{\tfrac{1}{\lambda}-\delta(1-\epsilon)}{\epsilon} \\
	0 & \textrm{otherwise}.
	\end{cases}
	\end{align*}
\end{cor}
\begin{IEEEproof}
	Similar to the proof of Cor.~\ref{cor:G/M/1_best}, it is sufficient to show that $\sigma^*$ is maximized, i.e., when $P^A$ is cramming $A^*(\mu(1-\sigma))$ upper bounds all other curves. We know that for any $P^A$,
	\begin{equation} \label{eq:worst_arrival}
	\begin{aligned}
	A^*(\mu(1-\sigma)) &= \int_{0}^{\infty} P^A(t) e^{-\mu t(1-\sigma)} dt \\
	&\le \int_{0}^{\infty} P^A(t) dt = 1.
	\end{aligned}
	\end{equation}
	On the other hand, note that the cramming inter-arrival distribution asymptotically achieves the upper bound as $\epsilon, \delta \to 0$ so that it maximizes the fixed point solution $\sigma^*$. Also notice that the location of $\epsilon$ point mass is determined to satisfy mean constraint $\mathbb{E}[A] = \lambda^{-1}$.
\end{IEEEproof}

\begin{cor}
	Fix service rate $\mu$. For $\M/\GI/1$ queues with channel quality stepping down at $b=0$, i.e.,
	\begin{align*}
		I(P_X^*, W_0) > I(P_X^*, W_1) = I(P_X^*, W_2) = \cdots,
	\end{align*}
	the capacity is constant among all service distributions.
\end{cor}
\begin{IEEEproof}
	When the threshold $b=0$, let $c_0 := I(P_X^*, W_0)$ and $c_1 := I(P_X^*, W_1)$. Since the capacity is given by
	\begin{align*}
		C(\Phi) = \pi(0) c_0 + (1-\pi(0)) c_1 = c_1 + \pi(0)(c_0 - c_1),
	\end{align*}
	so $\pi(0)$ completely determines the capacity. On the other hand, by the inverse $Z$-transform relation,
	\begin{align*}
		\pi(0) = \Pi(0) = 1-\rho.
	\end{align*}
	Thus, the capacity is constant over all $P^S$ of service rate $\mu$.
\end{IEEEproof}

\begin{cor} \label{cor:M/G/1_extrema}
	Fix service rate $\mu$. For $\M/\GI/1$ queues with channel quality stepping down at $b=1$, i.e.,
	\begin{align*}
		I(P_X^*, W_0) = I(P_X^*, W_1) > I(P_X^*, W_2) = \cdots,
	\end{align*}
	the capacity is maximized when the service is deterministic. On the other hand, the capacity is asymptotically minimized by cramming service, i.e., $P^S(t;\epsilon,\delta)$ asymptotically minimizes the capacity as $\epsilon,\delta \to 0$, where
	\begin{align*}
		P^S(t;\epsilon,\delta) = \begin{cases}
		1-\epsilon & \textrm{if } t=\delta \\
		\epsilon & \textrm{if } t = \frac{\frac{1}{\mu} - (1-\epsilon)\delta}{\epsilon} \\
		0 & \textrm{otherwise.}
		\end{cases}
	\end{align*}
\end{cor}
\begin{IEEEproof}
	Let $c_0:=I(P_X^*, W_0), c_2:=I(P_X^*, W_2)$ for simplicity. Then the capacity is given by
	\begin{align*}
		C = (\pi(0) + \pi(1)) c_0 + (1 - \pi(0) - \pi(1)) c_2.
	\end{align*}
	Since $c_0 > c_2$, it is apparent that the capacity is maximized (resp. minimized) when $\pi(0) + \pi(1)$ is maximized (resp. minimized). Also note that
	\begin{align*}
		\pi(0) &= 1-\rho = 1 - \frac{\lambda}{\mu}, \\
		\pi(1) &= \frac{\Pi(z)-\pi(0)}{z} \bigg|_{z=0} = \frac{ \frac{(1-\rho)(1-z)K(z)}{K(z)-z} - \pi(0)}{z} \bigg|_{z=0} \\
		&= \frac{ \frac{(1-\rho)(1-z)K(z)}{K(z)-z} - (1-\rho)}{z} \bigg|_{z=0} \\
		&= \frac{(1-\rho)(1-K(z))}{K(z)-z} \bigg|_{z=0} = \frac{(1-\rho) (1-K(0))}{K(0)}.
	\end{align*}
	Since $\pi(0) + \pi(1) = \frac{1-\rho}{K(0)}$, the best (resp. the worst) service distribution should minimize (resp. maximize) $K(0)=k_0$. Recall the expression of $k_0$,
	\begin{align*}
		k_0 = \int_{0}^{\infty} P^S(t) e^{-\lambda t} dt.
	\end{align*}
	The same arguments of \eqref{eq:best_arrival} and \eqref{eq:worst_arrival} imply that the deterministic service distribution $P^S(t) = \delta_{\mu^{-1}}$ maximizes the capacity, and 
	\begin{align*}
		P^S(t;\epsilon,\delta) = \begin{cases}
		1-\epsilon & \textrm{if } t=\delta \\
		\epsilon & \textrm{if } t = \frac{\frac{1}{\mu} - (1-\epsilon)\delta}{\epsilon} \\
		0 & \textrm{otherwise.}
		\end{cases}
	\end{align*}
	asymptotically minimizes the capacity as $\epsilon,\delta \to 0$.
\end{IEEEproof}
Cor.~\ref{cor:M/G/1_extrema} is also of interest when the number of users is large and each arrival process is sparse, as will be seen in Sec.~\ref{sec:RMPP_Queueing}.

\section{Multiuser Input: $\sum_k \GI_k/\GI/1$ Queues} \label{sec:multi-user}
Let $K,k$ be the total number of users and a specific $k$th user respectively, and recall the system model in Sec.~\ref{sec:system_model}. Since $K$ users simultaneously dispatch encoded symbols, each user sees a different queue-length distribution from that for a single-user system; thus, capacity changes. We characterize the individual and sum capacities, thus capacity region, for the $K$-user scenario in terms of $\pi_{Kk}(Q)$, the stationary queue-length distribution seen by user $k$'s arrivals. Since the superposition process is in general intractable, we obtain asymptotics of capacity using Poisson approximation when component PPs are independent and sparse.

For a common setup, consider a triangular array of independent, stationary, and renewal (thus, ergodic) PPs $\Phi_{Kk}, K \in \mathbb{Z}_+, k \in [1:K]$. Also suppose each PP has a continuous inter-arrival distribution $P_{Kk}^A$ with arrival rate $\lambda_{Kk}$, not necessarily identical. Let us also assume finite second-moment of inter-arrival times, which is necessary to prove Lem.~\ref{lem:indiv_ergodic}:
\begin{align}
	\mathbb{E}_{P_{Kk}^A}[A^2] < \infty \mbox{ for all } k \in [1:K]. \label{eq:arrival_moment_condi}
\end{align}

\subsection{Coding Theorem for $K$-user Channels} \label{sec:multi_fixed_K}
Let $\Phi_K$ be the superposition arrival process of $K$th-row components, i.e, $\Phi_K := \sum_{k=1}^K \Phi_{Kk}$. Note that the component PPs are stationary and ergodic.

The next lemma proves the superposition process is stationary and ergodic as well.
\begin{lem} \label{lem:ergodicity_superPP}
	Suppose each $\Phi_{Kk}, k \in [1:K]$ is independent, stationary, and ergodic. Then, $\Phi_K$ is also stationary and ergodic.
\end{lem}
\begin{IEEEproof}
	First prove the stationarity. Take an arbitrary bounded Borel set $B$ and let $B'=\mathcal{T}_{\tau}B$ be the time-shifted set by $\tau \in \mathbb{R}$. Consider the counting measure representation; then due to independence, $N_K(B) = \sum_k N_{Kk}(B)$ and
	\begin{align*}
		N_K(B) = \sum_k N_{Kk}(B) \stackrel{(a)}{=} \sum_k N_{Kk}(B') \stackrel{(b)}{=} N_K(B'),
	\end{align*}
	where (a) is due to the stationarity of individual PPs and (b) is due to independence of individual PPs. As $\tau \in \mathbb{R}$ is arbitrary, stationarity is shown.
	
	Next show the ergodicity. Suppose $\Phi_K$ is not ergodic: then, by Def.~\ref{def:ergodic_PP}, there exists a $Z \in \sigma(\mathcal{N})$ such that for any $\phi_K \in Z$ and $\tau \in \mathbb{R}$, it holds that $\mathcal{T}_{\tau} \phi_K \in Z$, however, $0 < P_K[Z] < 1$. As $Z$ is closed under any time-shift operation, we can write for $\phi_K \in Z$,
	\begin{align}
		&\phi_K = \left( \sum_k \phi_{Kk} \right) \in Z \Leftrightarrow \nonumber \\
		&\phi_K' := \mathcal{T}_{\tau} \phi_K = \mathcal{T}_{\tau} \sum_{k} \phi_{Kk} = \left( \sum_{k} \mathcal{T}_{\tau} \phi_{Kk} \right) \in Z ~ \forall \tau \in \mathbb{R}. \label{eq:Z_closedness}
	\end{align}
	
	Now consider $P_K[Z]$. Let $Z_k$ be the collection of $\phi_{Kk}$ consisting some $\phi \in Z$. As $\phi_{Kk}$ is a component of $\phi_K$, $\mathcal{T}_{\tau} \phi_{Kk}$ is also a component of $\phi_K'$ by \eqref{eq:Z_closedness} so that $Z_k$ is also closed. Since each $\Phi_{Kk}$ is stationary and ergodic, $P_{Kk}[Z_k]$ is either $0$ or $1$. However, because $0 < P_K[Z] = \prod_k P_{Kk}[Z_k] < 1$ by independence, there is a contradiction. Therefore, $\Phi_K$ is ergodic.
\end{IEEEproof}

Let $Q_i^{(K)}$ be the queue-length process seen by the superposed arrivals. The next lemma further guarantees that the stationary distribution $\pi_K$ exists and $Q_i^{(K)}$ is ergodic since $\Phi_K$ is stationary and ergodic from Lem.~\ref{lem:ergodicity_superPP}.
\begin{lem}[Chap.~2 \cite{FrankenKAS1982}] \label{lem:ergodicity_Q_response}
	If the input PP $\Phi$ of the queue $\cdot /\GI /1$ with traffic intensity $\rho < 1$ is stationary and ergodic, then the queue-length distribution seen by arrivals is also stationary and ergodic. Furthermore, the stationary distribution is independent of the initial state.
\end{lem}

Now let us consider individual `seen by arrivals' processes. Let $Q_i^{(Kk)}$ and $\pi_{Kk}$ denote the queue-length process seen by user $k$'s arrivals and its stationary distribution, respectively. The following lemma proves the existence of $\pi_{Kk}$ and its ergodicity.
\begin{lem} \label{lem:indiv_ergodic}
	Suppose \eqref{eq:arrival_moment_condi} holds. Then, for each $k \in [1:K]$, the stationary distribution $\pi_{Kk}$ exists. Furthermore, for any measurable $f:\mathbb{Z}_+ \mapsto \mathbb{R}_+$, $\frac{1}{n} \sum_{i=1}^n f(Q_i^{(Kk)}) \to \mathbb{E}_{\pi_{Kk}} [f(Q)]$ as $n \to \infty$ almost surely.
\end{lem}
\begin{IEEEproof}
	See App.~\ref{sec:app_ind_ergodic}.
\end{IEEEproof}

As before, Lem.~\ref{lem:indiv_ergodic} allows a simpler capacity expression. Let $C_{\textsf{ind}}(\Phi_{Kk}), C_{\textsf{sum}}(\Phi_K)$ be the $k$th user's individual capacity and their sum capacity. The following theorem only describes per job capacity, but per time capacity is immediate by multiplying by individual and sum arrival rates, respectively.
\begin{thm} \label{thm:individual_cap}
	\begin{align*}
		C_{\textsf{ind}}(\Phi_{Kk}) &= \mathbb{E}_{\pi_{Kk}} [I(P_X,W_Q)] ~~~ \textrm{[bits/sym]}, \\
		C_{\textsf{sum}}(\Phi_K) &= \mathbb{E}_{\pi_{K}} [I(P_X,W_Q)] = \sum_{k=1}^K w_k C_{\textsf{ind}}(\Phi_{Kk}) ~~ \textrm{[bits/sym]},
	\end{align*}
	where $w_k := \lambda_{Kk} / \sum_j \lambda_{Kj}$.
\end{thm}
\begin{IEEEproof}
	Since individual $\{\pi_{Kk}\}$ are stationary and ergodic, the first statement follows.
	
	To show the second statement, notice that
	\begin{align*}
		C_{\textsf{sum}}(\Phi_K) \le \mathbb{E}_{\pi_{K}} [I(P_X,W_Q)]
	\end{align*}
	holds. In addition, since $\pi_K$ is the weighted average of $\pi_{Kk}$, i.e., $\pi_K(q) = \sum_{k} w_k \pi_{Kk}(q)$, the equality holds.
\end{IEEEproof}

Unlike standard multiple-access settings, it is noteworthy that the per time sum capacity is simply the sum of per time individual capacities, which means that greedy individuals do not degrade optimality in sum information rate. This follows since once arrival processes are fixed, symbol noise levels are also fixed by the queue-length. The server processes one symbol at a time, therefore, adding more (or reducing) information in a user's codeword does not increase (or decrease) interference levels.

\subsection{Poisson Approximation} \label{sec:Poi_Approx}
In the previous subsection, we obtained the multiple-access capacity formula for general $\sum_{k} \GI/\GI/1$ queues. However, a more explicit expression is unavailable even for an $|\mathbb{F}|$-ary symmetric channel or an erasure channel, unless the queue is $\sum_{k} \M/\GI/1$. This is because the superposition of $K$ independent renewal PPs is not necessarily renewal and is renewal if and only if individual PPs are Poisson \cite{Samuels1974} (thus, the superposition process is also Poisson). So the superposition process is intractable. However, when $K$ is large and individual PPs are \emph{sparse} (formally defined in Def.~\ref{def:sparsity} below) we can approximate the superposition process by a Poisson PP.

Consider a triangular array of i.i.d., stationary, ergodic, and renewal PPs, $\{\Phi_{Kk}\}$, where $K \in \mathbb{Z}_+$ and $k \in [1:K]$. Individual processes are assumed to be sparse as given below. The superposition process of row PPs is denoted by $\Phi_K := \sum_{k} \Phi_{Kk}$ with corresponding probability measure $P_K$. Let $N_{Kk}(B)$ be the counting measure corresponding to $\Phi_{Kk}$, i.e., the number of events of $\Phi_{Kk}(t)$ in $B \in \mathfrak{B}$. Also let $N_K(B)$ be the number of events of $\Phi_K$ in $B$, so $N_K(B) = \sum_{k} N_{Kk}(B)$. We can then derive that $N_K(B)$ converges to the Poisson distribution of intensity measure $\lambda |B|$ where $|\cdot|$ is the Lebesgue measure, or equivalently, $\Phi_K(t)$ converges to the Poisson process, say $\Phi^*(t)$, with probability measure $P^*$, under the sparsity condition. Let $N^*$ be the counting measure for the Poisson PP, i.e., for any bounded $B \in \mathfrak{B}$,
\begin{align*}
	\Prob[N^*(B)=j] = \frac{1}{j!} (\lambda|B|)^j e^{-\lambda |B|}.
\end{align*}

\begin{defi}
\label{def:sparsity}
For a given bounded $B \in \mathfrak{B}$, the triangular processes are said to be \textit{sparse with sum rate} $\lambda_K := \sum_k \lambda_{Kk} + \frac{g_1(K,B)}{|B|}$ if 
\begin{itemize}
	\item $\lambda_{Kk} := \frac{\Prob[N_{Kk}(B) = 1]}{|B|}$
	
	\item $g_1(K,B) := \sum_{k=1}^K \sum_{j=2}^{\infty} j \Prob[N_{Kk}(B) = j] \to 0$ as $K \to \infty$
	
	\item $g_2(K) := \max_{k \in [1:K]} \lambda_{Kk} \to 0$ as $K \to \infty$
	
	\item $g_2(K)$ converges sufficiently fast so that for a sequence of bounded intervals $B = B(\epsilon_K) \in \mathfrak{B}$ with $\epsilon_K \downarrow 0$, $|B(\epsilon_K)|^2 g_2(K) \to 0$ holds.
\end{itemize}
\end{defi}
The first three conditions ensure that the superposition process converges to the Poisson process and the last condition further ensures the convergence of the queue-length in Lem.~\ref{lem:q_length_TV}.

The next lemma shows that $\Phi_K$ \textit{locally} converges to $\Phi^*$ on $B$ in total variation sense. The lemma holds for any bounded $B \in \mathfrak{B}$, but we focus on a bounded interval $B=[a,b]$. The proof basically relies on so-called Poisson approximation, available in various forms, e.g., \cite{Grigelionis1963, Kallenberg2017}, but here we provide convergence result in total variational form with explicit convergence bound. Let $\lambda_K^* := \sum_k \lambda_{Kk}$.
\begin{lem}\label{lem:Poisson_approx}
	Fix a bounded $B \in \mathfrak{B}$ of interest and let $\Phi_K^*$ be the Poisson PP with intensity $\lambda_K^* |B|$. Suppose individual PPs of the triangular array are sparse with sum rate $\lambda_K$. Then, $N_K(B) \to N_K^*(B)$ in total variation. Furthermore, the speed of convergence is $O(g(K,B))$, where $g(K,B) := \max \{ g_1(K,B), |B|^2 g_2(K)\}$.
\end{lem}
\begin{IEEEproof}
	See App.~\ref{sec:app_poi_approx}.
\end{IEEEproof}

The next corollary is especially useful in the next subsection, where each user sends symbols on i.i.d.\ renewal arrivals.

\begin{cor} \label{cor:iid_component_conv}
	Suppose component PPs in a row of the triangular array are identically distributed, and $\lambda_K^* = \lambda$ for all $K$, i.e., Poisson PPs corresponding to each row are identical. Then, $d_{\textsf{TV}}(N_K(B),N^*(B)) \to 0$ as $K \to \infty$ with speed $O(g_1(K,B), |B|^2 K^{-1})$, where $N^*$ is the counting measure for the Poisson PP with intensity $\lambda$.
\end{cor}

\subsection{Capacity Approximation} \label{sec:RMPP_Queueing}
We reformulate input processes of the queue as two-sided RMPPs to streamline proofs and arguments. Recall that the mark space $\mathcal{M}=\mathbb{R}_+$ and the fact that service times are drawn i.i.d.~from $P^S$. Suppose that the RMPPs begin at $t=-T$ for large $T>0$ and the queue is initially empty. Since all randomness of queueing is captured by the RMPP, any queue-state process is a deterministic function of $\Phi(t)$ and initial queue state $\theta_{-T}$, which is empty. For example, discrete-time queue state processes, such as a queue-length process seen by arrival or departure, can be expressed as $z(i, \Phi, \theta_{-T})$ for some deterministic function $z$. As we have seen previously, the process of queue-length seen by arrivals $\{Q_i\}_{i \in \mathbb{Z}}$ is of interest. Note that
\begin{align*}
	Q_i(\Phi) = h(i, \Phi, \theta_{-T}) ~ \textrm{ for some deterministic function } h.
\end{align*}

Consider the case of Cor.~\ref{cor:iid_component_conv}, where users' individual arrivals are i.i.d.\ and corresponding Poisson sum rate is $\lambda_K^* = \lambda$ for all $K$. As corresponding Poisson PPs are identically distributed regardless of $K$, we drop the subscript $K$ if clear from the context. Let $Q_i^{(K)}$ be the queue-length process seen by $i$th arrival of the $K$-user superposition process. Similarly let $Q_i^*$ be the corresponding process for the Poisson PP $\Phi^* (=\Phi_K^* \mbox{ for all } K)$. Then, the continuity theorem holds due to the locally convergence property above. Here, $\stackrel{\textsf{TV}}{\to}$ denotes local convergence of PP on $B \in \mathfrak{B}$ in total variation. For random variables, $\stackrel{\textsf{TV}}{\to}$ is the usual total variational convergence.

\begin{lem} \label{lem:q_length_TV}
	For any $\epsilon > 0$, we can take $K_0 \in \mathbb{Z}_+$ and an interval $B = B(\epsilon) \in \mathfrak{B}$ that yields
	\begin{align*}
		d_{\textsf{TV}}(Q_i^{(K)}, Q_i^*) \le 2 \epsilon + O(g(K,B)), \quad K \ge K_0
	\end{align*}
	where $g(K,B) = \max \{ g_1(K,B), |B|^2 g_2(K)\}$. Therefore, the last condition of Def.~\ref{def:sparsity} provides $Q_i^{(K)} \stackrel{\textsf{TV}}{\to} Q_i^*$.
\end{lem}
\begin{IEEEproof}
	See App.~\ref{sec:app_q_length_TV}.
\end{IEEEproof}

Recall notations that $\pi_{Kk}, \pi_{K}$ denote the stationary queue-length distributions seen by an individual user's and superposed arrivals, respectively. As individual users are symmetric, $\pi_{Kk}$ are identical and in addition $\pi_{Kk} = \pi_K$ for all $k$.

Since each arrival has only a few arrivals on $B$ (with high probability), we implicitly suppose the transmission is repeated many times to achieve block code performance.

Let $c_{\textsf{max}}:=\sup_q \max_{P_X} I(P_X,W_q)$, which is $c_{\textsf{max}} \le \log |\mathcal{X}|$ clearly. The final approximation, our main result, follows.
\begin{thm} \label{thm:sum_approx_error}
	Let $C(\Phi^*)$ be the single-user capacity of $\M/\GI/1$ queue with arrival rate $\lambda$, derived in Cor.~\ref{thm:M/G/1_queue}. Consider $K$ users with sparse individual PPs $\Phi_{Kk}$. Then, under superposition, the sum capacity $C_{\textsf{sum}}(\Phi_{K})$ at arrival rate $\lambda$ is approximated by the single-user capacity $C(\Phi^*)$ as
	\begin{align*}
		& | C_{\textsf{sum}} (\Phi_{K}) - C(\Phi^*) | \le c_{\textsf{max}} \left( 4\epsilon + O(g(K, B)) \right) ~ \textrm{[bits/sym]}, \\
		&| C_{\textsf{sum}} (\Phi_{K}) - C(\Phi^*) | \\
		&\le \frac{g_1(K,B)}{|B|} c_{\textsf{max}} + \lambda c_{\textsf{max}} \left( 4\epsilon + O(g(K, B)) \right) ~ \textrm{[bits/time]}.
	\end{align*}
\end{thm}
\begin{IEEEproof}
	As $\pi_K = \pi_{Kk}$ for all $k$, individuals can send information at rate 
	\begin{align*}
		C(\Phi_{Kk}) = \sum_{q} \pi_{Kk}(q) I(P_X, W_q) ~~~ \textrm{[bits/sym]}, 
	\end{align*}
	the sum rate is also $C(\Phi_{Kk})$ in bits per symbol sense. On the other hand, the stationary distribution $\pi_K$ differs from the stationary distribution for Poisson, say $\pi^*$, at most $2\epsilon + O(g(K, B))$ in total variation. This implies
	\begin{align*}
		& | C_{\textsf{sum}} (\Phi_{K}) - C(\Phi^*) | \\
		&= \left| \sum_{q=0}^{\infty} (\pi^*(q) - \pi_{K}(q)) I(P_X, W_q) \right| \\
		&\le c_{\textsf{max}} \left| \sum_{q=0}^{\infty} (\pi^*(q) - \pi_{K}(q)) \right| \\
		&\le c_{\textsf{max}} \sum_{q=0}^{\infty} \left|\pi^*(q) - \pi_{K}(q) \right| = c_{\textsf{max}} \cdot 2 d_{\textsf{TV}}(Q_k^{(K)}, Q^*) \\
		&\le c_{\textsf{max}}(4\epsilon + O(g(K,B))).
	\end{align*}
	
	To obtain the second bound, recall that actual sum arrival rate of the superposition process deviates from $\lambda$ by $\frac{g_1(k,B)}{|B|}$. Therefore,
	\begin{align*}
		&| C_{\textsf{sum}} (\Phi_{K}) - C(\Phi^*) | \\
		&= \Big\lvert \left( \lambda + \frac{g_1(K,B)}{|B|} \right) \sum_q \pi_K(q) I(P_X, W_q) \\
		&~~~~ ~~~~ - \lambda \sum_q \pi^*(q) I(P_X, W_q) \Big\rvert \\
		&\le \frac{g_1(K,B)}{|B|} c_{\textsf{max}} + \lambda c_{\textsf{max}} \cdot 2 d_{\textsf{TV}}(Q_k^{(K)}, Q^*) \\
		&\le \frac{g_1(K,B)}{|B|} c_{\textsf{max}} + \lambda c_{\textsf{max}} \left( 4\epsilon + O(g(K, B)) \right) ~~ \textrm{[bits/time]}
	\end{align*}
\end{IEEEproof}

Thm.~\ref{thm:sum_approx_error} only considers the sum capacity, however, it is clear from the proof that individual per symbol capacity remains unchanged, and per time capacity is properly scaled, i.e.,
\begin{align*}
	& \left| C_{\textsf{ind}}(\Phi_{Kk}) - \frac{C(\Phi^*)}{K} \right| \\
	&\le \frac{g_1(K,B)}{K|B|} c_{\textsf{max}} + \frac{\lambda}{K} c_{\textsf{max}} \left( 4\epsilon + O(g(K, B)) \right) ~ \textrm{[bits/time]}.
\end{align*}
Therefore, the best and worst server results in Cor.~\ref{cor:M/G/1_extrema} also apply to the superposition arrivals asymptotically as $K \to \infty$.
\begin{cor}
	Suppose the conditions in Sec.~\ref{subsec:optimization_cap} hold. Then, for the $K$-user setting with sparse individuals, the results in Cor.~\ref{cor:M/G/1_extrema} still hold asymptotically, that is, when the channel quality steps down at $b=1$, the sum and individual capacities are maximized when the service is deterministic. On the other hand, the sum and individual capacities are asymptotically minimized by cramming service.
\end{cor}

\section{Conclusion} \label{sec:conclusion}
In this paper, we have presented the capacity of the queue-length dependent channel in a multiple-access setting. We modelled the workload and its buffering process as a queueing process with noise and characterized the capacity of single-user and multiple-access systems. We first obtain the capacity in multi-letter form, however, the ergodicity of the queue enables us to derive single-letter expressions in Thms.~\ref{thm:single_user_cap} and \ref{thm:individual_cap}. Unlike standard multiple-access problems, information rate in codewords does not change other users' performance as in Thm.~\ref{thm:individual_cap}. Furthermore, when the number of users is large and each arrival process is sparse, the individual and sum capacities are asymptotically close to the single-user capacity of $\M/\GI/1$ queues, and thus, the best (resp. the worst) service in single-user is also the best (resp. the worst) in multiple-access.

Since Shannon capacity can often only be attained with long decoding delay from a large block length, this paper provides a system design benchmark in case of delay-sensitive systems such as multimedia communication, finance markets, or autonomous driving. More refined analysis can be performed using the finite block length technique \cite{Polyanskiy2010} or appropriate coding schemes that guarantee a finite delay.

\appendices
\section{Proof of Thm.~\ref{thm:single_user_cap}} \label{sec:app_pf_thm1}
In Prop.~\ref{prop:coding_thm}, we have the capacity expression
\begin{align*}
	C(\Phi) = \sup_{P_{\mathbf{X}}} \underline{\mathbf{I}}(\mathbf{X};\mathbf{Y}|\mathbf{Q}),
\end{align*}
which is in infinite-letter form.

Notice that the supremum optimizes over joint distributions $P_{\mathbf{X}}(X^n) = P_{\mathbf{X}}(X_1, X_2, \ldots, X_n)$. We can further bound the expression by a supremum expression over product distribution space as follows.

Let $\overline{\mathbf{X}} = (\overline{X}_1, \overline{X}_2, \ldots, \overline{X}_n)$ be a sequence of random variables whose probability distribution is in product form
\begin{align*}
P_{ \overline{\mathbf{X}}}(X^n) = P_{\mathbf{X}}(X_1) P_{\mathbf{X}}(X_2) \cdots P_{\mathbf{X}}(X_n).
\end{align*}
That is, the product of marginals of $P_{\mathbf{X}}$. The induced received symbols $\overline{\mathbf{Y}}$ are similarly defined via channels. Using \cite[Lem.~3.2.3]{Han2003}, we have the following inequality.
\begin{align*}
\underline{\mathbf{I}}(\mathbf{X};\mathbf{Y}|\mathbf{Q}) \le \underline{\mathbf{I}}(\overline{\mathbf{X}};\overline{\mathbf{Y}}|\mathbf{Q}) := \liminf_{n \to \infty} \frac{1}{n} \sum_{i=1}^n I(P_{\mathbf{X}}(X_i); W_{Q_i}).
\end{align*}
Therefore, we have an upper bound which is a function of a product distribution.

Let $I_{q} = \{i \in \mathbb{Z}_+ | Q_i = q\}$, that is, the time instances when $Q_i = q$. Then,
\begin{align*}
&\frac{1}{n} \sum_{i=1}^n I(P_{\mathbf{X}}(X_i); W_{Q_i}) \\
&= \frac{1}{n} \left[ \sum_{i \in I_0} I(P_{\mathbf{X}}(X_i); W_{0}) + \sum_{i \in I_1} I(P_{\mathbf{X}}(X_i); W_{1}) + \cdots \right] \\
&= \sum_{q=0}^{\infty} \frac{|I_q|}{n} \cdot \frac{1}{|I_q|} \sum_{i \in I_q} I(P_{\mathbf{X}}(X_i); W_{q}) \\
&\le \sum_{q=0}^{\infty} \frac{|I_q|}{n} \sum_{i \in I_q} I(P_{X, q} ; W_{q}),
\end{align*}
where the inequality follows from the convexity of mutual information with $P_{X,q}$ being the average of $P_{\mathbf{X}}(X_i)$ over $i \in I_q$. Since the encoder has no access to the queue-length realization, $P_{X,q} = P_X$ for all $q$, and thus, $I(P_{X, q} ; W_{q}) = I(P_X;W_q)$. The ergodic property in Lem.~\ref{lem:stationary_ergodic_lem} yields $|I_q|/n \to \pi(q)$ so we have
\begin{align*}
\underline{\mathbf{I}}(\mathbf{X};\mathbf{Y}|\mathbf{Q}) \le \sum_{q=0}^{\infty} \pi(q) I(P_X, W_q).
\end{align*}
As this bound is attainable by taking $P_{\mathbf{X}}(X^n) = \prod_i P_{X}(X_i)$, the claim has been proved.

\section{Proof of Lem.~\ref{lem:indiv_ergodic}} \label{sec:app_ind_ergodic}
To prove the `seen by arrival' result, we start from arbitrary-time ergodicity in \cite{DaiM1995}. We first take a continuous-time piecewise-deterministic Markov process \cite{Davis1984}. Then, since it is strong Markov, the stopped process at user $k$ arrivals forms a stationary and ergodic discrete-time Markov chain. Suppose that once job processing is completed and the job departs at time $t$, the next job enters the server at time $t^+$.

Let us take a continuous-time Markov process $\mathbf{Z}(t) := (\mathbf{L}(t), \mathbf{A}(t), \mathbf{S}(t)) \in \mathcal{Z}$, where
\begin{itemize}
	\item $\mathbf{L}(t)$ is the vector of user indices of jobs in the system in order of their arrivals including the job in the server. If the system is empty, $\mathbf{L}(t) = \emptyset$. Otherwise, $\mathbf{L}(t) = (\ell_1, \ell_2, \ldots, \ell_{Q(t)}) \in [1:K]^{Q(t)}$, where $Q(t)$ is the queue-length at time $t$, and each $\ell_i$ indicates user index of $i$th job in the system. $\ell_1$ is the user index of the job being served.
	
	\item $\mathbf{A}(t) \in \mathbb{R}_+^K$ is the residual arrival time vector whose component $A_k(t)$ indicates the remaining time until the
	next arrival of $k$th user.
	
	\item $\mathbf{S}(t) \in (\mathbb{R}_+ \cup \infty)^K$ is the residual service time vector whose component $S_k(t)$ indicates residual service time if user $k$'s job is being served, infinite otherwise.
\end{itemize}
Under condition \eqref{eq:arrival_moment_condi}, this is Harris recurrent so there exists an arbitrary-time stationary distribution $\hat{\pi}$ and the following holds \cite[Thm.~6.4]{DaiM1995}: For any $g: \mathcal{Z} \mapsto \mathbb{R}_+$,
\begin{align}
	\lim_{t \to \infty} \frac{1}{t} \int_{0}^t g(\mathbf{Z}(s)) ds = \mathbb{E}_{\hat{\pi}}[g(\mathbf{Z})] \mbox{ almost surely}. \label{eq:DaiM_ergodicity}
\end{align}

Fix a user $k$ and take a sequence of stopping times $(t_1, t_2, \ldots)$ such that $t_n := \min \{t > t_{n-1}: A_k(t-) > 0, A_k(t) = 0\}$ (assume $t_0 < 0$ for simplicity), i.e., the sequence of hitting times at which user $k$th job arrives. Take a small $\Delta > 0$ and two indicators $g_1 := \mathbf{1}_{\{A_k(t) \le \Delta\}}, g_2 := \mathbf{1}_{\{|\mathbf{L}(t)| = q, A_k(t) \le \Delta\}}$. Since inter-arrival time distributions are Lebesgue continuous, \eqref{eq:DaiM_ergodicity} implies
\begin{align*}
	&\lim_{n \to \infty} \frac{1}{t_n} \int_{0}^{t_n} g_1(\mathbf{Z}(s)) ds \\
	&\qquad = \Delta \cdot \hat{\pi}\{\mathbf{Z}(t):A_k(t) = 0\} + O(\Delta^2), \\
	&\lim_{n \to \infty} \frac{1}{t_n} \int_{0}^{t_n} g_2(\mathbf{Z}(s)) ds \\
	&\qquad = \Delta \cdot \hat{\pi}\{\mathbf{Z}(t):Q(t) = q, A_k(t) = 0\} + O(\Delta^2).
\end{align*}
Taking $\Delta \to 0$ and using the fact that the queue-length is a deterministic function of $\mathbf{L}(t)$, it follows that the stationary distribution exists and
\begin{align}
	\pi_{Kk}(q) := \frac{\hat{\pi}\{\mathbf{Z}(t):|\mathbf{L}(t)| = q, A_k(t) = 0\}}{\hat{\pi}\{\mathbf{Z}(t):A_k(t) = 0\}}. \label{eq:stat_dist_from_conti}
\end{align}

Next show the ergodicity. Define two samplings
\begin{align*}
	h_1(\mathbf{Z}(t)) &:= \mathbf{1}_{ \{ A_k(t) \le \Delta \} }, \\
	h_2(\mathbf{Z}(t)) &:= \mathbf{1}_{ \{ A_k(t) \le \Delta \} } f(q(t)),
\end{align*}
and note that
\begin{align*}
	\lim_{n \to \infty} \frac{1}{t_n} \int_{0}^{t_n} h_1(\mathbf{Z}(s))ds = \lim_{n \to \infty} \frac{n \Delta}{t_n} = \lambda_{Kk} \Delta
\end{align*}
and 
\begin{align*}
	& \lim_{n \to \infty} \frac{1}{t_n} \int_{0}^{t_n} h_2(\mathbf{Z}(s))ds = \lim_{n \to \infty} \frac{1}{t_n} \sum_{i=1}^n f(q(t_j)) \Delta \\
	&= \lim_{n \to \infty} \frac{n}{t_n} \frac{1}{n} \sum_{i=1}^n f(q(t_i)) \Delta = \lambda_{Kk} \Delta \lim_{n \to \infty} \frac{1}{n} \sum_{i=1}^n f(q(t_i)),
\end{align*}
where $\lim_n \frac{n}{t_n} \to \lambda_{Kk}$ is assumed due to the system stability. Then,
\begin{align}
	\frac{\lim_{n \to \infty} \frac{1}{t_n} \int_{0}^{t_n} h_2(\mathbf{Z}(s)) ds }{\lim_{n \to \infty} \frac{1}{t_n} \int_{0}^{t_n} h_1(\mathbf{Z}(s)) ds} = \lim_{n \to \infty} \frac{1}{n} \sum_{i=1}^n f(q(t_i)). \label{eq:sampling}
\end{align}

Also letting $\Delta \to 0$ and applying \eqref{eq:DaiM_ergodicity} to the left side of \eqref{eq:sampling}, 
\begin{align}
	& \frac{\lim_{n \to \infty} \frac{1}{t_n} \int_{0}^{t_n} h_2(\mathbf{Z}(s)) ds }{\lim_{n \to \infty} \frac{1}{t_n} \int_{0}^{t_n} h_1(\mathbf{Z}(s)) ds} = \frac{ \mathbb{E}_{\hat{\pi}}[h_2(\mathbf{Z})] }{ \mathbb{E}_{\hat{\pi}}[h_1(\mathbf{Z})] } \nonumber \\
	&= \frac{\sum_{q=0}^{\infty} f(q) \hat{\pi} \{ \mathbf{Z}(t): A_k(t) = 0, |\mathbf{L}(t)| = q \}}{\hat{\pi}\{\mathbf{Z}(t): A_k(t) = 0\}} \nonumber \\
	&= \sum_{q=0}^{\infty} f(q) \frac{ \hat{\pi} \{ A_k(t) = 0, |\mathbf{L}(t)| = q \}}{\hat{\pi}\{A_k(t) = 0\}} \nonumber \\
	&= \sum_{q=0}^{\infty} f(q) \pi_{Kk}(q) = \mathbb{E}_{\pi_{Kk}}[f(Q)]. \label{eq:ergodic_last}
\end{align}
Since $Q_i = Q(t_i)$, the following holds from \eqref{eq:sampling} and \eqref{eq:ergodic_last},
\begin{align*}
	\lim_{n \to \infty} \frac{1}{n} \sum_{i=1}^n f(q_i) = \lim_{n \to \infty} \frac{1}{n} \sum_{i=1}^n f(q(t_i)) = \mathbb{E}_{\pi_{Kk}}[f(Q)]
\end{align*}
almost surely.

\section{Proof of Lem.~\ref{lem:Poisson_approx}} \label{sec:app_poi_approx}
We restricted to PPs on a bounded $B$ so $\Phi_K, \Phi_K^*$ both have no events outside of $B$. Therefore it is sufficient to show that for all $B' \in \mathfrak{B}$ such that $B' \subset B$,
\begin{align*}
	d_{\textsf{TV}}( N_K(B'), N_K^*(B')) \to 0 \textrm{ as } K \to \infty.
\end{align*}

Note that Poisson processes are infinitely divisible, so we can split into $K$ independent Poisson PPs $\{\Phi_{Kk}^*\}_{k \in [1:K]}$ with intensity $\lambda_{Kk}$. Let $N_{Kk}^*$ be the counting measure of $\Phi_{Kk}^*$. From the Poisson distribution and its Taylor expansion when $|B|\lambda_{Kk}$ is small:
\begin{align*}
	\Prob[N_{Kk}^*(B) = 1] &= |B|\lambda_{Kk} + O(|B|^2\lambda_{Kk}^2), \\
	\Prob[N_{Kk}^*(B) \ge 2] &= O(|B|^2\lambda_{Kk}^2).
\end{align*}
Hence, total variational distance between individual PPs is computed as follows, where argument $B$ is omitted for simplicity.
\begin{align*}
	& 2 d_{\textsf{TV}}(N_{Kk}, N_{Kk}^*) \\
	&= \sum_{j \in \mathbb{Z}_+} \Big| \Prob[N_{Kk}=j] - \Prob[N_{Kk}^*=j] \Big| \\
	&= \Big| (1-\Prob[N_{Kk} \ge 1]) - (1-\Prob[N_{Kk}^* \ge 1]) \Big| \\
	& ~~~~ + \sum_{j \ge 1} \Big| \Prob[N_{Kk}=j] - \Prob[N_{Kk}^*=j] \Big| \\
	&= \Big| \Prob[N_{Kk}^* = 1] + \Prob[N_{Kk}^* \ge 2] - \Prob[N_{Kk} = 1] \\
	& ~~~~ - \Prob[N_{Kk} \ge 2] \Big| + \sum_{j \ge 1} \Big| \Prob[N_{Kk}=j] - \Prob[N_{Kk}^*=j] \Big| \\
	&\stackrel{(a)}{\le} \Big| \Prob[N_{Kk}^* = 1] - \Prob[N_{Kk} = 1] \Big| + \Prob[N_{Kk}^* \ge 2] \\
	& ~~~~ + \Prob[N_{Kk} \ge 2] + \sum_{j \ge 1} \Big| \Prob[N_{Kk}=j] - \Prob[N_{Kk}^*=j] \Big| \\
	&\stackrel{(b)}{\le} \Big| \Prob[N_{Kk}^* = 1] - |B|\lambda_{Kk} \Big| + O(|B|^2\lambda_{Kk}^2) + \Prob[N_{Kk} \ge 2] \\
	& ~~~ + \sum_{j \ge 1} \Big| \Prob[N_{Kk}=j] - \Prob[N_{Kk}^*=j] \Big| \\
	&\stackrel{(c)}{\le} O(|B|^2\lambda_{Kk}^2) + \Prob[N_{Kk} \ge 2] \\
	& ~~~ + \sum_{j \ge 1} \Big| \Prob[N_{Kk}=j] - \Prob[N_{Kk}^*=j] \Big| \\
	&\stackrel{(d)}{\le} O(|B|^2\lambda_{Kk}^2) + \Prob[N_{Kk} \ge 2] + \Prob[N_{Kk} \ge 2] + \Prob[N_{Kk}^* \ge 2] \\
	&\stackrel{(e)}{=} O(|B|^2\lambda_{Kk}^2) + 2 \Prob[N_{Kk} \ge 2], 
\end{align*}
where (a) follows from the triangle inequality; (b) follows from the first condition of Def.~\ref{def:sparsity} and the Taylor expansion; (c) follows from the Taylor expansion; (d) follows from the triangle inequality; and (e) follows from the Taylor expansion.

Now we bound total variation between two sums of independent random variables as follows.
\begin{align*}
	& d_{\textsf{TV}}(N_{K}, N_K^*) \\
	&\stackrel{(a)}{\le} \sum_{k \in [1:K]} d_{\textsf{TV}}( N_{Kk}, N_{Kk}^* ) \\
	&\stackrel{(b)}{\le} \sum_{k \in [1:K]} O\left( |B|^2 \lambda_{Kk}^2 \right) + \sum_{k \in [1:K]} \Prob[N_{Kk} \ge 2] \\
	&\le c |B|^2 \cdot \sum_{k \in [1:K]} \lambda_{Kk} \left( \max_{k \in [1:K]} \lambda_{Kk} \right)  + \sum_{k \in [1:K]} \Prob[N_{Kk} \ge 2] \\
	&= c |B|^2 \cdot \lambda_K^* \cdot g_2(K) + \sum_{k \in [1:K]} \Prob[N_{Kk} \ge 2],
\end{align*}
where (a) follows from the total variation inequality for product measures, and (b) follows from the above derivation.	

Therefore, the first term vanishes at speed $O(|B|^2g_2(K))$, the second term $\sum_{k} P[N_{Kk} \ge 2] \to 0$ at speed $O(g_1(K,B))$. So the overall speed of convergence is given by $O(g(K,B))$, where $g(K,B) := \max \{g_1(K,B), |B|^2g_2(K) \}$.

Finally, for all subsets $B' \subset B$ with $B' \in \mathfrak{B}$, we can repeat the above argument, but the speed of convergence still holds since $g_1(K,B') \le g_1(K,B)$ and $|B'|g_2(K) \le |B|g_2(K)$.

\section{Proof of Lem.~\ref{lem:q_length_TV}} \label{sec:app_q_length_TV}
We will first restrict the superposed RMPP on some $B$, and then apply the data processing inequality (also known as monotone theorem in some literature \cite{Reiss1993}) to show $Q_i^{(K)} \stackrel{\textsf{TV}}{\to} Q_i^*$. Without loss of generality, we only consider some arbitrary $i$th symbol whose arrival was at $t_i>0$.

Let us introduce an \emph{empty point} \cite{FrankenKAS1982}. When $\phi(t)$ is a specific realization of $\Phi(t)$, an arrival time instance $e_j(\phi)$ at which there is no job in the system (in the queue and in the server both) is called an empty point.\footnote{This is different from the regenerative cycles, introduced in Sec.~\ref{sec:single-user}. Since we are considering arbitrary superposition process $\Phi$ that is not renewal in general, so $e_j(\Phi)$ is not regenerative.} List $e_j(\phi)$ in order 
\begin{align*}
	\cdots < e_{-1}(\phi) < e_0(\phi) \le 0 < e_1(\phi) < \cdots.
\end{align*}
The $j$th empty point implies that the queue state after $t=e_j(\phi)$ is completely determined only by arrivals after $e_j(\phi)$.
Then, we know that $e_0(\Phi_K) \stackrel{\textsf{TV}}{\to} e_0(\Phi^*)$ with speed $O(g(K,B))$ by data processing inequality and thus, $e_j(\Phi_K) \stackrel{\textsf{TV}}{\to} e_j(\Phi^*)$ for any $j$ by stationarity.

Take a set of PP realizations $A_{u_1} := \{\phi: -u_1< e_0(\phi) \le 0 \}$. Since $e_0(\Phi_K) \stackrel{\textsf{TV}}{\to} e_0(\Phi^*)$, for arbitrary $\epsilon_1 > 0$ it is possible to take $u_1, K_0$ such that for all $K > K_0$,
\begin{align*}
	P_K[A_{u_1}] > 1- \epsilon_1 \textrm{ and } P^*[A_{u_1}] > 1-\epsilon_1.
\end{align*}
Also, take a set $A_{u_2}^i := \{\phi: 0 < t_i(\phi) < u_2 \}$. Thus it is immediate that for arbitrary $\epsilon_2 > 0$ we can take $u_2>0$ such that $P_K[A_{u_2}^i] > 1- \epsilon_2 ~ \textrm{ and } ~ P^*[A_{u_2}^i] > 1-\epsilon_2$.

Let $q(i, \phi)$ be the queue-length seen by $i$th arrival of $\phi$, and $u:=\max(u_1, u_2), \epsilon:=\epsilon_1 + \epsilon_2$. By the property of the empty point and $A_{u_1}, A_{u_2}^i$,
\begin{align*}
	P^*[\phi: q(i, \phi) &= q(i, \mathbf{1}_{[-u,u)} \phi)] \ge P^*[A_{u_1} \cap A_{u_2}^i] > 1-\epsilon, \\
	P_K[\phi: q(i, \phi) &= q(i, \mathbf{1}_{[-u,u)} \phi)] \ge P_K[A_{u_1} \cap A_{u_2}^i] > 1-\epsilon.
\end{align*}

Setting $B = [-u, u)$, we can bound total variation as follows.
\begin{align*}
	& d_{\textsf{TV}}( Q_i(\Phi_K), Q_i(\Phi^*) ) \\
	&\stackrel{(a)}{\le} d_{\textsf{TV}}( Q_i(\Phi_K), Q_i(\mathbf{1}_{B} \Phi_K)) + d_{\textsf{TV}}( Q_i(\mathbf{1}_{B} \Phi_K), Q_i(\mathbf{1}_{B} \Phi^*)) \\
	&~~~ + d_{\textsf{TV}}( Q_i(\mathbf{1}_B \Phi^*), Q_i(\Phi^*)) \\
	&\stackrel{(b)}{\le} 2\epsilon + d_{\textsf{TV}}( Q_i(\mathbf{1}_{B} \Phi_K), Q_i(\mathbf{1}_{B} \Phi^*)) \\
	&\stackrel{(c)}{\le} 2\epsilon + d_{\textsf{TV}}( \mathbf{1}_{B} \Phi_K, \mathbf{1}_{B} \Phi^*) \\
	&\stackrel{(d)}{\le} 2\epsilon + O(g(K,B)).
\end{align*}
where (a) follows from the triangle inequality; (b) follows from the property of empty point; (c) follows from the data processing inequality since $Q_i(\cdot)$ is a function of a PP; and (d) follows from Lem.~\ref{lem:Poisson_approx}. Since $\epsilon_1, \epsilon_2$ are arbitrary, the statement is proved. The last condition of Def.~\ref{def:sparsity} provides $Q_i^{(K)} \stackrel{\textsf{TV}}{\to} Q_i^*$.


\begin{thebibliography}{99}
	
	\bibitem{SeoCV2018}
	D.~Seo, A.~Chatterjee, and L.~R. Varshney, ``On multiuser systems with
	queue-length dependent service quality,'' in \emph{Proc. 2018 IEEE Int. Symp.
		Inf. Theory}, Jun. 2018, pp. 341--345.
	
	\bibitem{Aldridge2013}
	I.~Aldridge, \emph{High-Frequency Trading: A Practical Guide to Algorithmic
		Strategies and Trading Systems}, 2nd~ed.\hskip 1em plus 0.5em minus
	0.4em\relax Hoboken, NJ: John Wiley \& Sons, 2013.
	
	\bibitem{PapadimitratosFEBC2009}
	P.~Papadimitratos, A.~D.~L. Fortelle, K.~Evenssen, R.~Brignolo, and S.~Cosenza,
	``Vehicular communication systems: Enabling technologies, applications, and
	future outlook on intelligent transportation,'' \emph{{IEEE} Commun. Mag.},
	vol.~47, no.~11, pp. 84--95, Nov. 2009.
	
	\bibitem{SriramL1989}
	K.~Sriram and D.~M. Lucantoni, ``Traffic smoothing effects of bit dropping in a
	packet voice multiplexer,'' \emph{{IEEE} Trans. Commun.}, vol.~37, no.~7, pp.
	703--712, Jul. 1989.
	
	\bibitem{DraperTW2005}
	S.~C. Draper, M.~D. Trott, and G.~W. Wornell, ``A universal approach to queuing
	with distortion control,'' \emph{{IEEE} Trans. Autom. Control}, vol.~50,
	no.~4, pp. 532--537, Apr. 2005.
	
	\bibitem{Goyal2001b}
	V.~K. Goyal, ``Multiple description coding: Compression meets the network,''
	\emph{{IEEE} Signal Process. Mag.}, vol.~18, no.~5, pp. 74--93, Sep. 2001.
	
	\bibitem{Schwartz1978}
	B.~Schwartz, ``Queues, priorities, and social process,'' \emph{Soc. Psychol.},
	vol.~41, no.~1, pp. 3--12, Mar. 1978.
	
	\bibitem{DugdaleEP1999}
	D.~C. Dugdale, R.~Epstein, and S.~Z. Pantilat, ``Time and the patient-physician
	relationship,'' \emph{J. Gen. Intern. Med.}, vol.~14, no.~S1, pp. S34--S40,
	Jan. 1999.
	
	\bibitem{DerletR2000}
	R.~W. Derlet and J.~R. Richards, ``Overcrowding in the nation's emergency
	departments: Complex causes and disturbing effects,'' \emph{Ann. Emerg.
		Med.}, vol.~35, no.~1, pp. 63--68, Jan. 2000.
	
	\bibitem{Jamal2007}
	M.~Jamal, ``Job stress and job performance controversy revisited: An empirical
	examination in two countries,'' \emph{Int. J. Stress Management}, vol.~14,
	no.~2, pp. 175--187, May 2007.
	
	\bibitem{Higginbotham2019}
	S.~Higginbotham, ``Autonomous trucks need people,'' \emph{{IEEE} Spectr.},
	vol.~56, no.~3, p.~21, Mar. 2019.
	
	\bibitem{ChatterjeeSV2017}
	A.~Chatterjee, D.~Seo, and L.~R. Varshney, ``Capacity of systems with
	queue-length dependent service quality,'' \emph{{IEEE} Trans. Inf. Theory},
	vol.~63, no.~6, pp. 3950--3963, Jun. 2017.
	
	\bibitem{FrankenKAS1982}
	P.~Franken, D.~K{\"o}nig, U.~Arndt, and V.~Schmidt, \emph{Queues and Point
		Processes}.\hskip 1em plus 0.5em minus 0.4em\relax New York: John Wiley \&
	Sons, 1982.
	
	\bibitem{DaleyV1998}
	D.~J. Daley and D.~Vere-Jones, \emph{An Introduction to the Theory of Point
		Processes}.\hskip 1em plus 0.5em minus 0.4em\relax Berlin, Germany:
	Springer-Verlag, 1998.
	
	\bibitem{Grigelionis1963}
	B.~Grigelionis, ``On the convergence of sums of random step processes to a
	{P}oisson process,'' \emph{Theory Probab. Appl.}, vol.~8, no.~2, pp.
	177--182, Jun. 1963.
	
	\bibitem{Kallenberg2017}
	O.~Kallenberg, \emph{Random Measures, Theory and Applications}.\hskip 1em plus
	0.5em minus 0.4em\relax Cham, Switzerland: Springer, 2017.
	
	\bibitem{HouBK2009}
	I.-H. Hou, V.~Borkar, and P.~R. Kumar, ``A theory of {QoS} for wireless,'' in
	\emph{Proc. 2009 IEEE INFOCOM}, Apr. 2009.
	
	\bibitem{HouK2010}
	I.-H. Hou and P.~R. Kumar, ``Utility-optimal scheduling in time-varying
	wireless networks with delay constraints,'' in \emph{Proc. 11th ACM Int.
		Symp. Mobile ad hoc Networking and Computing}, Sep. 2010, pp. 31--40.
	
	\bibitem{JaramilloS2010}
	J.~J. Jaramillo and R.~Srikant, ``Optimal scheduling for fair resource
	allocation in ad hoc networks with elastic and inelastic traffic,'' in
	\emph{Proc. 2010 IEEE INFOCOM}, Mar. 2010.
	
	\bibitem{KaulYG2012}
	S.~Kaul, R.~Yates, and M.~Gruteser, ``Real-time status: How often should one
	update?'' in \emph{Proc. 2012 IEEE INFOCOM}, Mar. 2012.
	
	\bibitem{CostaCE2014}
	M.~Costa, M.~Codreanu, and A.~Ephremides, ``Age of information with packet
	management,'' in \emph{Proc. 2014 IEEE Int. Symp. Inf. Theory}, Jun. 2014,
	pp. 1583--1587.
	
	\bibitem{KamKNE2015}
	C.~Kam, S.~Kompella, G.~D. Nguyen, and A.~Ephremides, ``Effect of message
	transmission path diversity on status age,'' \emph{{IEEE} Trans. Inf.
		Theory}, vol.~62, no.~3, pp. 1360--1374, Mar. 2015.
	
	\bibitem{SunUYKS2017}
	Y.~Sun, E.~Uysal-Biyikoglu, R.~D. Yates, C.~E. Koksal, and N.~B. Shroff,
	``Update or wait: How to keep your data fresh,'' \emph{{IEEE} Trans. Inf.
		Theory}, vol.~63, no.~11, pp. 7492--7508, Nov. 2017.
	
	\bibitem{TelatarG1995}
	{\.{I}}.~E. Telatar and R.~G. Gallager, ``Combining queueing theory with
	information theory for multiaccess,'' \emph{{IEEE} J. Sel. Areas Commun.},
	vol.~13, no.~6, pp. 963--969, Aug. 1995.
	
	\bibitem{RajTT2004}
	S.~Raj, E.~Telatar, and D.~Tse, ``Job scheduling and multiple access,'' in
	\emph{Advances in Network Information Theory}, P.~Gupta, G.~Kramer, and A.~J.
	van Wijngaarden, Eds.\hskip 1em plus 0.5em minus 0.4em\relax Providence:
	DIMACS, American Mathematical Society, 2004, pp. 127--137.
	
	\bibitem{ChenCG2017}
	X.~Chen, T.-Y. Chen, and D.~Guo, ``Capacity of {G}aussian many-access
	channels,'' \emph{{IEEE} Trans. Inf. Theory}, vol.~63, no.~6, pp. 3516--3539,
	Jun. 2017.
	
	\bibitem{AnantharamV1996}
	V.~Anantharam and S.~{Verd\'{u}}, ``Bits through queues,'' \emph{{IEEE} Trans.
		Inf. Theory}, vol.~42, no.~1, pp. 4--18, Jan. 1996.
	
	\bibitem{BedekarA1998}
	A.~S. Bedekar and M.~Azizo{\~{g}}lu, ``The information-theoretic capacity of
	discrete-time queues,'' \emph{{IEEE} Trans. Inf. Theory}, vol.~44, no.~2, pp.
	446--461, Mar. 1998.
	
	\bibitem{Ahlswede1971}
	R.~Ahlswede, ``Multi–way communication channels,'' in \emph{Proc. 2nd Int.
		Symp. Inf. Theory}, Sep. 1971, pp. 103--135.
	
	\bibitem{Ahlswede1974}
	------, ``The capacity region of a channel with two senders and two
	receivers,'' \emph{Ann. Probab.}, vol.~2, pp. 805--814, Oct. 1974.
	
	\bibitem{VerduH1994}
	S.~{Verd\'{u}} and T.~S. Han, ``A general formula for channel capacity,''
	\emph{{IEEE} Trans. Inf. Theory}, vol.~40, no.~4, pp. 1147--1157, Jul. 1994.
	
	\bibitem{Han2003}
	T.~S. Han, \emph{Information-Spectrum Methods in Information Theory}.\hskip 1em
	plus 0.5em minus 0.4em\relax Berlin: Springer, 2003.
	
	\bibitem{CoverT1991}
	T.~M. Cover and J.~A. Thomas, \emph{Elements of Information Theory}.\hskip 1em
	plus 0.5em minus 0.4em\relax New York: John Wiley \& Sons, 1991.
	
	\bibitem{Asmussen2003}
	S.~Asmussen, \emph{Applied Probability and Queues, 2nd ed.}\hskip 1em plus
	0.5em minus 0.4em\relax New York, USA: Springer-Verlag, 2003.
	
	\bibitem{Kleinrock1975_I}
	L.~Kleinrock, \emph{Queuing Systems, Volume I: Theory}.\hskip 1em plus 0.5em
	minus 0.4em\relax John Wiley \& Sons, Inc., 1975.
	
	\bibitem{Samuels1974}
	S.~M. Samuels, ``A characterization of the {P}oisson process,'' \emph{J. Appl.
		Probab.}, no.~1, pp. 72--85, Mar. 1974.
	
	\bibitem{Polyanskiy2010}
	Y.~Polyanskiy, ``Channel coding: non-asymptotic fundamental limits,'' Ph.D.
	dissertation, Princeton University, Nov. 2010.
	
	\bibitem{DaiM1995}
	J.~Dai and S.~Meyn, ``Stability and convergence of moments for multiclass
	queueing networks via fluid limit models,'' \emph{{IEEE} Trans. Autom.
		Control}, vol.~40, no.~11, pp. 1889--1904, Nov. 1995.
	
	\bibitem{Davis1984}
	M.~H.~A. Davis, ``Piecewise-deterministic {M}arkov processes: A general class
	of non-diffusion stochastic models,'' \emph{J. Roy. Stat. Soc. Ser. B},
	vol.~46, no.~3, pp. 353--388, 1984.
	
	\bibitem{Reiss1993}
	R.-D. Reiss, \emph{A Course on Point Processes}.\hskip 1em plus 0.5em minus
	0.4em\relax New York, USA: Springer--Verlag, 1993.
	
\end{thebibliography}
\end{document}